\documentclass{elsart}
\usepackage{latexsym,amsmath,tabularx,amssymb,bm}
\usepackage{epsfig}
\usepackage{citesort}

\long\def\comment#1{ }

\usepackage{graphicx}
\usepackage{amssymb}
\usepackage{makeidx}
\usepackage{graphicx}
\usepackage{multicol}

\def\simge{\mathrel{%
    \rlap{\raise 0.511ex \hbox{$>$}}{\lower 0.511ex \hbox{$\sim$}}}}
\def\simle{\mathrel{
    \rlap{\raise 0.511ex \hbox{$<$}}{\lower 0.511ex \hbox{$\sim$}}}}

\newcommand \beq{\begin{eqnarray}}
\newcommand \eeq{\end{eqnarray}}
\newcommand{\del}{\partial}

\begin{document}
\begin{flushright}
~\vspace{-1.25cm}\\
{\small\sf ECT*-- 05-02}
\end{flushright}
\vspace{0.8cm}
\begin{frontmatter}

\parbox[]{16.6cm}{ \begin{center}

\title{A new method to solve the Non Perturbative Renormalization Group equations}

\author[ect]{J.-P. Blaizot \thanksref{cnrs}},
\author[montevideo]{Ram\'on M\'endez Galain\thanksref{email2}}, 
\author[montevideo]{Nicol\'as Wschebor\thanksref{email}}

\address[ect]{$ECT*$, Villa Tambosi, Strada delle Tabarelle 286,
I-38050 Villazzano(TN), Italy}

\address[montevideo]{ Instituto de F\'{\i}sica, Facultad de Ingenier\'{\i}a,
J.H.y Reissig 565, 11000
  Montevideo, Uruguay}

\thanks[cnrs]{Membre du Centre National de la Recherche Scientifique
(CNRS), France.}

\thanks[email2]{email: mendezg@fing.edu.uy}

\thanks[email]{email: nicws@fing.edu.uy}

\date{\today}
\vspace{0.8cm}
\begin{abstract}
We propose a method to solve the Non Perturbative Renormalization Group equations
 for the $n$-point functions. In leading order, it 
consists in solving the equations obtained by closing the infinite
hierarchy of equations for the $n$-point functions. This is
achieved: i) by exploiting the decoupling of modes and the analyticity of the $n$-point functions at small momenta: this allows us
to neglect some momentum dependence of the vertices entering the
flow equations; ii) by relating vertices at zero momenta to
derivatives of lower order vertices with respect to a constant
background field. Although the approximation is not controlled by
a small parameter, its accuracy can be systematically improved.
When it is applied to the $O(N)$ model, its leading order is exact  in
the large $N$ limit; in this case, one
recovers known results in a simple and direct way, i.e., without introducing
an auxiliary field.

\end{abstract}
\end{center}}

\end{frontmatter}
\newpage

\section{Introduction}

In the last ten years a considerable amount of effort has been 
devoted to the study of the Non Perturbative Renormalization Group 
(NPRG), and to the development of new approximation schemes to solve 
the corresponding infinite hierarchy of equations for the $n$-point 
functions \cite{Tetradis94,Ellwanger94a,Morris94b,Morris94c}.  
In this context, the  so-called ``derivative expansion" has been 
widely used. It defines a systematic approach that has led to 
many successful applications  in a variety of domains 
\cite{Berges00,Bagnuls:2000ae,Delamotte:2004zg}. However,  the
derivative expansion is a good approximation to $n$-point
functions only when the external momenta are smaller than the
lowest mass in the problem. In particular, for massless theories,
the derivative expansion provides information only on the
$n$-point functions (and their derivatives) at zero momenta.  
This is not enough for applications which require the knowledge 
of the full momentum dependence of the $n$-point functions. In these cases, 
a new approximation scheme is necessary.

To our knowledge, all efforts in this direction \cite{truncation}
have been based on various forms
 of the early proposal by Weinberg \cite{weinberg73}: 
one {\it truncates} the infinite tower of flow equations for the $n$-point 
functions considering only vertices up to a given number of legs, 
possibly using various ansatzs for some of them. 
This leads to approximations similar to those
used when solving Schwinger-Dyson equations \cite{alkofer}.
However, despite the fact that very encouraging results have been obtained 
problems remain with truncations around zero
external fields \cite{convergence}.

It is then interesting to look for alternative schemes to solve NPRG at finite 
external momenta and it is
 the purpose of  this Letter  to present a  method to calculate  correlation functions at arbitrary
momenta within the NPRG. As it is the case for the derivative expansion,
the proposed strategy takes into account, 
at each order, an infinite number of vertices.
The  accuracy is expected to be comparable to that achieved with the 
derivative expansion when it is used to calculate the effective potential.
The method exploits two properties of the NPRG:  the decoupling of 
short wave-length modes, and the
analyticity of the $n$-point functions at small momentum
(guaranteed by the infrared regulator), in order to neglect some
of the momentum dependence in the vertices which enter the flow
equations. Then, it becomes possible to close the infinite
hierarchy of equations by calculating vertices at zero momenta as
derivatives of lower order vertices with respect to a constant
background field. Although the approximation is not controlled by
any small parameter, its accuracy can be systematically improved, in 
a way similar to what can be done in the derivative expansion.

The leading order of the proposed approximation scheme shares some of
the nice properties of the leading order of the derivative
expansion, the so-called local potential approximation (LPA): it
is exact at one loop order,  and it also reproduces the exact
large $N$ limit of the $O(N)$ model. Of course, the LPA provides
only information on momentum independent quantities, such as the
effective potential, whereas the present method does not have this
limitation.  In fact, as an illustration, we use it  to solve  the NPRG equations for
the full momentum dependent $n$-point functions of the $O(N)$
model in the large $N$ limit. This can be done directly, i.e.,  
without having to introduce an auxiliary
field. To our knowledge, this is the first time that, within the NPRG, 
such a calculation is done for this model.

The method presented here is an improvement of  the one  that we have applied in a
previous paper to the calculation of the transition temperature of a weakly repulsive Bose gas
\cite{Blaizot:2004qa}. Compared to the method proposed in
\cite{Blaizot:2004qa}, the present one  is conceptually simpler as it
involves a single approximation. Its numerical implementation  will be discussed in a
separate publication \cite{BMW05_3}.

\section{The NPRG and the derivative expansion}

Let us start by recalling some basic features of  the NPRG. Although most of the arguments in
this paper have a wider range of applicability, we shall consider a scalar field
 theory defined by an Euclidean action $S$ in $d$ dimensions, of
the generic form:
\begin{equation}\label{eactON}
S = \int {\rm d}^{d}x\,\left\lbrace{ 1 \over 2}   \left(\del_\mu
\varphi(x)\right)^2  +V(\varphi)\right\rbrace \,,
\end{equation} where $V(\varphi)$ is a polynomial in $\varphi$. It
is understood that the parameters of the action  (hidden in
$V(\varphi)$) and the field normalization are fixed at a
microscopic momentum scale $\Lambda$ (which may be infinite). The
NPRG equations   relate the classical action to the full effective
action. This relation is obtained by controlling the magnitude of
long wavelength field fluctuations with the help of an infrared
cut-off,  which is implemented
\cite{Tetradis94,Ellwanger94a,Morris94b,Morris94c} by adding to
the classical  action (\ref{eactON}) a regulator of the form
\begin{equation}
  \Delta S_\kappa[\varphi] =\frac{1}{2} \int \frac{{\rm d}^dp}{(2\pi)^d}
R_\kappa(p)
\varphi(p)\varphi(-p),
  \end{equation}\normalsize
where $R_\kappa$ denotes a family of ``cut-off functions''
depending on a parameter $\kappa$.  The role of $\Delta S_\kappa$
is to suppress the fluctuations with momenta $p\simle \kappa$,
while leaving unaffected the modes with $p\simge \kappa$. Thus,
typically $ R_{\kappa}(p)\to\kappa^2$ when $ p \ll \kappa$, and
$R_{\kappa}(p)\to 0$
 when $ p\simge \kappa$.
There is a large
freedom in the choice of $R_\kappa(p)$, abundantly discussed in the literature
\cite{Ball95,Comellas98,Litim,Canet02}.

We denote  the effective action in the presence of the regulator by
 $\Gamma_\kappa[\phi]$, where $\phi$ is the average field, $\phi(x)=\left\langle \varphi(x)
\right\rangle$. When $\kappa \to \Lambda$ quantum fluctuations are suppressed and $\Gamma_\Lambda[\phi]$ coincides with the classical action. As $\kappa$ decreases,
 more and more quantum fluctuations are taken into account and, as  $\kappa\to 0$, $\Gamma_{\kappa=0}[\phi]$ becomes the usual effective action $\Gamma[\phi]$. In other words, as
$\kappa$ decreases from $\Lambda$ to $0$,  $\Gamma_\kappa$
interpolates between the classical action and the full
effective action (see e.g. \cite{Berges00}). The variation
with $\kappa$ of $\Gamma_\kappa[\phi]$ is governed by the
following flow
equation \cite{Tetradis94,Ellwanger94a,Morris94b,Morris94c}:
\begin{equation}
\label{NPRGeq}
\partial_\kappa \Gamma_\kappa[\phi]=\frac{1}{2}\int \frac{d^dq}{(2\pi)^d} \partial_\kappa R_\kappa(q^2)
\left[\Gamma_\kappa^{(2)}+R_\kappa\right]^{-1}_{q,-q},
\end{equation}
where $\Gamma_\kappa^{(2)}$ is the second derivative of $\Gamma_\kappa$ w.r.t. $\phi$ (see Eq.~(\ref{gamma2}) below).
Eq.~\ref{NPRGeq} is the master equation of the NPRG. Its right
hand side has the structure of a one loop integral, with one
insertion of $\partial_\kappa R_\kappa(q^2)$.

\begin{figure}
\begin{center}
\includegraphics[scale=.5] {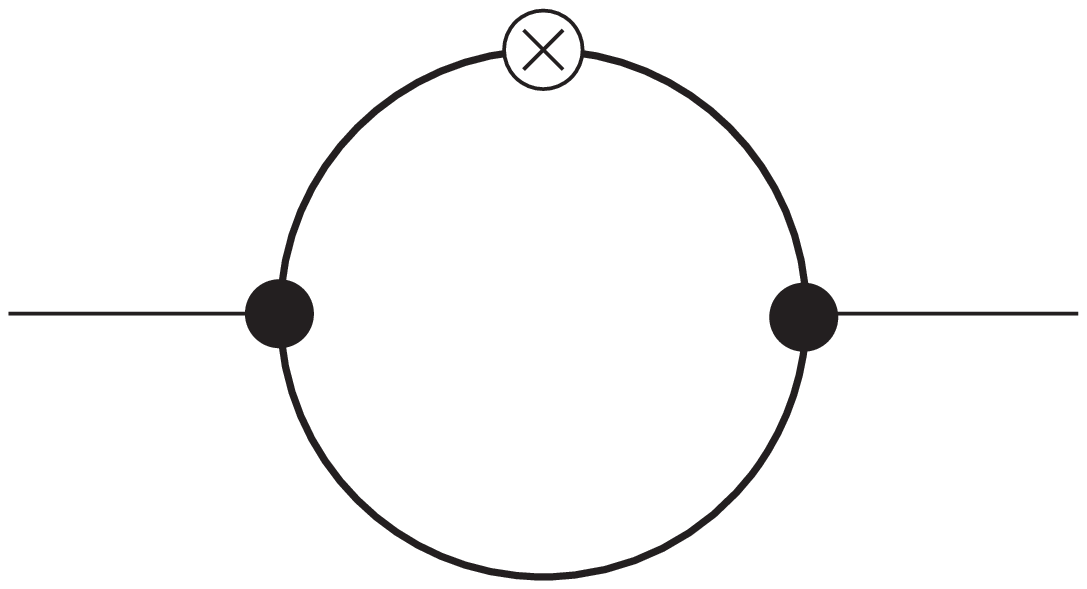} \hspace{20mm}
\includegraphics[scale=.5] {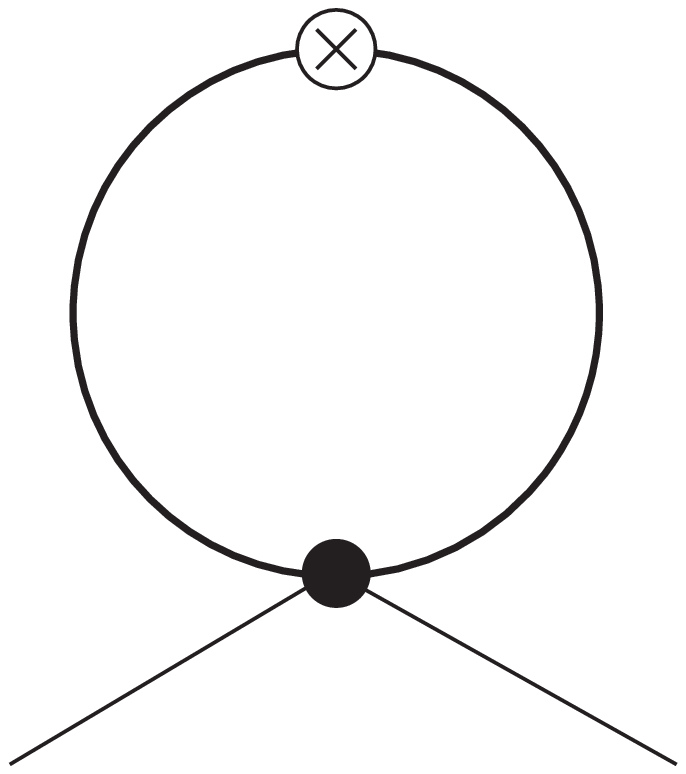}
\end{center}
\caption{The two diagrams contributing to the flow of the 2-point vertex. The lines represent dressed propagators, $G_\kappa$. The
 cross represents an insertion of $\partial_\kappa R_k$. The vertices denoted by   black dots are $\Gamma^{(3)}_\kappa$ and $\Gamma^{(4)}_\kappa$. \label{2-point-diagrams}}
\end{figure}

 As well known \cite{Zinn-Justin:2002ru}, the effective action $\Gamma[\phi]$ is the generating functional
of the one-particle irreducible $n$-point functions. Similarly, for a given value of $\kappa$, we define the $n$-point functions $\Gamma_\kappa^{(n)}$:
\begin{equation}
\Gamma_\kappa^{(n)}(p_1,\dots,p_n)=\int d^dx_1\dots\int d^dx_{n-1}
e^{i\sum_{j=1}^n p_jx_j}\frac{\delta^n\Gamma_\kappa}{\delta\phi(x_1)
\dots \delta\phi(x_n)}.
\end{equation}
By deriving eq.~(\ref{NPRGeq}) with respect to $\phi$, and then
letting the field be constant, one gets the flow equations for all $n$-point
functions in a constant background field $\phi$.  These equations can
be represented diagramaticaly by  one loop  diagrams
with dressed vertices and propagators (see e.g. \cite{Berges96}). For instance, the
flow of the 2-point
function in a constant external field reads:
\beq
\label{gamma2}
\partial_\kappa\Gamma_\kappa^{(2)}(p,-p;\phi)&=&\int
\frac{d^dq}{(2\pi)^d}\partial_\kappa R_k(q)\left\{G_\kappa(q^2;\phi)\Gamma_\kappa^{(3)}(p,q,-p-q;\phi)\right. \nonumber \\
&&\times G_\kappa((q+p)^2;\phi)\Gamma_\kappa^{(3)}(-p,p+q,-q;\phi)G_\kappa(q^2;\phi) \nonumber \\
&&\left.-\frac{1}{2}G_\kappa(q^2;\phi)\Gamma_\kappa^{(4)}(p,-p,q,-q;\phi)G_\kappa(q^2;\phi)\right\},
\eeq where
\begin{equation}\label{G-gamma2}
G^{-1}_\kappa (q^2;\phi) = \Gamma^{(2)}_\kappa (q,-q;\phi) +
R_\kappa(q^2).
\end{equation}The corresponding diagrams contributing to the flow are shown
in Fig.~\ref{2-point-diagrams}.

Flow equations for the $n$-point functions do not close: for example, in order to solve eq.~(\ref{gamma2}) one needs
the 3- and the 4-point functions, $\Gamma_\kappa^{(3)}$ and $\Gamma_\kappa^{(4)}$ respectively.  At this point we observe that because of the shape of the regulator, only internal momenta $q$ smaller than $\kappa$ contribute to the flow, i.e., to the integral in the r.h.s. of eq.~(\ref{gamma2}). One refers to this property as to the decoupling of high momentum modes. Besides,  the regulator insures also that all vertices  are smooth functions of momenta.
 Suppose then that one wants to calculate $n$-point
functions at small external momenta, say, $p_i^2 \simle
\kappa^2$. Then all momenta are small.
 This, together with the fact that the $n$-point functions are smooth functions of the momenta, make it
  possible to expand the $n$-point functions  $\Gamma_\kappa^{(n)}$ in the r.h.s. of the flow equations in terms of $q^2/\kappa^2$ and $p^2/\kappa^2$, or equivalently in terms of the derivatives of the field.

  Such considerations are at the basis of the  derivative expansion, which is usually formulated in terms of an ansatz for the effective action, rather than in terms of an approximation for the $n$-point functions.
  Its zero order, the LPA, assumes that the effective action has the form:
\begin{equation}\label{gammaLPA} \Gamma_k^{LPA}[\phi]=\int d^dx \left\{
\frac{1}{2}\partial_{\mu}\phi\partial_{\mu}\phi+V_k(\phi)\right\}.
\end{equation}
The derivative term here is simply the one appearing in the
classical action, and $V_k(\phi)$ is the effective potential. The
flow equation for $V_\kappa$ is a closed equation that is easily
obtained  by assuming that the field $\phi$ is constant in
eq.~(\ref{NPRGeq}):
\begin{equation}
\label{pot}
\partial_\kappa V_\kappa (\phi)=\frac{1}{2}\int \frac{d^dq}{(2\pi)^d} \partial_\kappa R_\kappa(q^2) \; G_\kappa^{LPA}(q^2;\phi),
\end{equation}
with $G_\kappa^{LPA}(q^2;\phi)$ given by Eq.~(\ref{G-gamma2}) in which 
\begin{equation}
\label{prop}
\Gamma_\kappa^{(2)}(q,-q;\phi)=q^2+\del ^2V_\kappa/\del\phi^2, 
\end{equation}
as obtained from Eq.~(\ref{gammaLPA})

Higher order corrections to the LPA include terms in the effective action with an
increasing number of derivatives. Although there is no formal
proof of convergence,
 the expansion exhibits quick apparent
convergence if the
regulator $R_\kappa(q^2)$ is appropriately chosen  \cite{Litim01,Canet02,Canet03}. In practice, the LPA reproduces well the physical quantities dominated by  small momenta (such as the effective potential or
critical exponents) in all theories where it has been tested (see,
for example, \cite{Berges00,Delamotte:2004zg}). Higher order corrections
improve the results. The expansion has been pushed
up to third order  \cite{Canet03}, yielding  critical exponents in
the Ising universality class as good as those obtained with the
best accepted methods.

\section{A new approximation scheme }

The   derivative expansion, as described above, strictly  makes sense  only for momenta
 not much larger than $\sqrt{\kappa^2+m_\kappa^2}$, where $m_\kappa$ is the running mass.  Thus, at criticality, and in the physical limit $\kappa \to 0 $, it provides  information only on $n$-point functions and their derivatives at zero momenta. However, by focusing on the  $n$-point functions rather than the effective
 action, we can generalize slightly  the arguments on which the derivative expansion is based in order to set up a much more powerful approximation scheme. We observe that:
 i)  the momentum $q$ circulating in the loop integral of a flow equation
   is limited by $\kappa$; ii)  the
   smoothness of the $n$-point functions allows us to make an expansion in powers of $q^2/\kappa^2$, independently of the value  of the external momenta $p$. Now, a typical $n$-point function entering a flow equation is of the from $\Gamma^{(n)}_\kappa(p_1,p_2,...,p_{n-1}+q,p_n-q;\phi)$, where $q$ is the loop momentum. The proposed approximation scheme, \emph{in its leading order},  will then consist in neglecting the  $q$-dependence of such vertex functions:
\begin{equation}\label{approx}
\Gamma^{(n)}_\kappa(p_1,p_2,...,p_{n-1}+q,p_n-q;\phi)\sim
\Gamma^{(n)}_\kappa(p_1,p_2,...,p_{n-1},p_n;\phi).
\end{equation}
Note that this approximation is a priori  well justified.
Indeed, when all the external momenta $p_i=0$, eq.~(\ref{approx})
is the basis of the LPA  which, as stated above, is a good
approximation. When the external momenta $p_i$ begin to grow, the
approximation in eq.~(\ref{approx}) becomes better and better,  and it is
trivial when all momenta are much larger than $\kappa$. With this
approximation,  eq.~(\ref{gamma2}) for instance becomes:
\begin{eqnarray}
\label{gamma2app}
\partial_\kappa\Gamma_\kappa^{(2)}(p,-p;\phi)&=&\int
\frac{d^dq}{(2\pi)^d}\partial_\kappa R_k(q^2)\left\{G_\kappa(q^2;\phi)\Gamma_\kappa^{(3)}(p,0,-p;\phi)\right. \nonumber \\
&&\times G_\kappa((q+p)^2;\phi)\Gamma_\kappa^{(3)}(-p,p,0;\phi)G_\kappa(q^2;\phi) \nonumber \\
&&\left.-\frac{1}{2}G_\kappa(q^2;\phi)\Gamma_\kappa^{(4)}(p,-p,0,0;\phi)G_\kappa(q^2;\phi)\right\}
\end{eqnarray}
Note that we do not also assume $q=0$ in the propagators. The reason for this will become clear shortly.

Now comes the second ingredient of the approximation scheme, which
exploits the advantage of working with a non vanishing background
field:  vertices evaluated at  zero external momenta can be
written  as derivatives of vertex functions with a smaller number
of legs (this observation can be found, in a  different
context, in \cite{Tetradis94} and in \cite{Golner:1998sr}):
\begin{equation}
\label{faireq=0}
\Gamma_\kappa^{(n+1)}(p_1,p_2,...,p_n,0;\phi)=\frac{\partial
\Gamma_\kappa^{(n)}(p_1,p_2,...p_n;\phi)} {\partial \phi}.
\end{equation}
To prove this relation,  expand $\Gamma_k[\phi]$ around an arbitrary  constant field
 $\phi^0$:
\begin{eqnarray}
\label{seriedelafonctionnelle} \Gamma_\kappa[\phi]&=&\sum_n
\frac{1}{n!}\int d^dx_1 ... d^dx_n\, [\phi(x_1)-\phi^0]...
[\phi(x_n)-\phi^0]\,\Gamma^{(n)}_\kappa(x_1,...,x_n;\phi_0) .
\end{eqnarray}
 Since $\Gamma_\kappa[\phi]$ does not depend on $\phi^0$, $\del\Gamma_\kappa[\phi]/\del \phi^0=0$.
Taking the derivative of eq.~(\ref{seriedelafonctionnelle}) with respect to $\phi_0$ one then gets:
\begin{equation} \int
d^dy\,
\Gamma_\kappa^{(n+1)}(x_1,x_2,...,x_n,y,\phi^0)=\frac{\partial
\Gamma_\kappa^{(n)}(x_1,x_2,...x_n,\phi^0)} {\partial \phi^0}
,\end{equation}from which eq.~(\ref{faireq=0}) follows after a
Fourier transform. By exploiting eq.~(\ref{faireq=0}), one easily
transforms eq.~(\ref{gamma2app}) into a {\it closed equation} (recall
that $G_\kappa$ and $\Gamma_\kappa^{(2)}$ are related by
eq.~(\ref{G-gamma2})):
\begin{eqnarray}\label{2pointclosed}
&&\partial_\kappa\Gamma_\kappa^{(2)}(p^2;\phi)=\int
\frac{d^dq}{(2\pi)^d}\,\partial_\kappa R_\kappa(q^2) \; G^2_\kappa(q^2;\phi)  \nonumber \\
&&
\times \left\{ \left( \frac{\partial \Gamma_\kappa^{(2)}(p,-p;\phi)}
{\partial \phi} \right)^2 G_\kappa ((p+q)^2;\phi)
\; - \; \frac{1}{2}\frac{\partial^2 \Gamma_\kappa^{(2)}(p,-p;\phi)}
{\partial \phi^2}\right\}.
\end{eqnarray}

It is interesting to emphasize the similarity of this equation with  
eq.~(\ref{pot}): both are closed equations because
the vertices appearing in the r.h.s. have been expressed
as derivatives of the function in the l.h.s.. 
This is the main result of this paper.

The approximation scheme presented here is similar to that used in \cite{Blaizot:2004qa}. There also the momentum dependence of the vertices was neglected in the leading order. However further approximations were needed in order to close the hierarchy. The progress realized here is to bypass these extra approximations by working in a constant background field.

\begin{figure}
\begin{center}
\includegraphics[scale=.5] {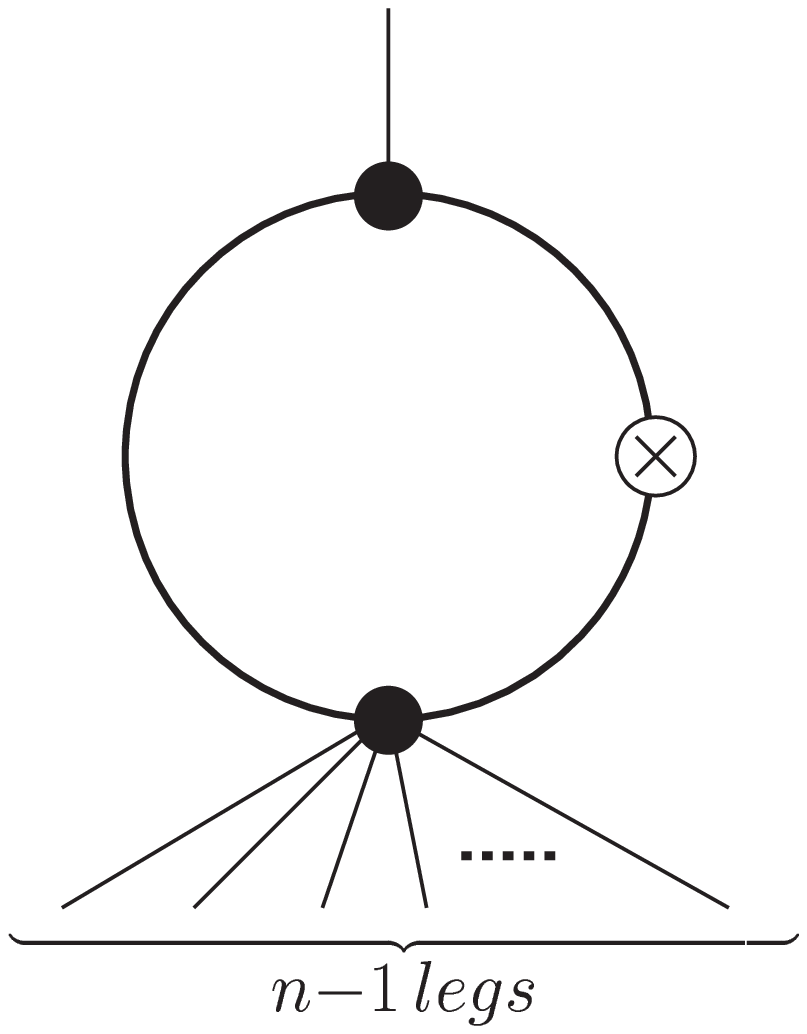}\hspace{25mm}
\includegraphics[scale=.5] {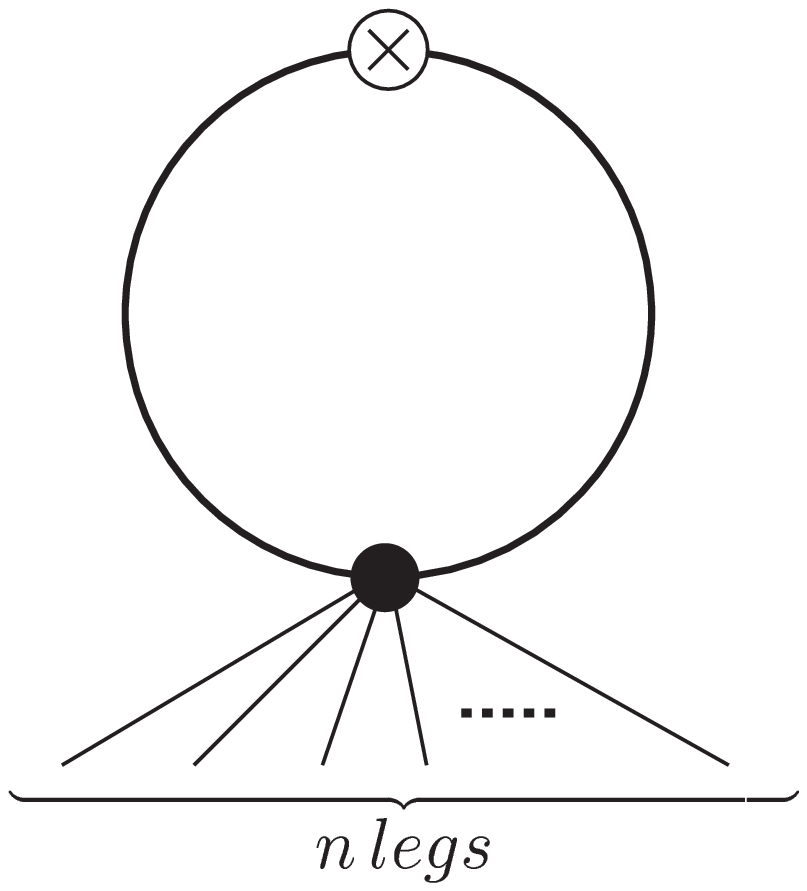}
\end{center}
\caption{The two diagrams contributing to the flow of the
$n$-point vertex and containing vertices with more than $n$ legs.
\label{n-point-diagrams}}
\end{figure}

The construction of closed equations for the $n$-point
functions with arbitrary $n$  $ (n \geq 3$) follows the same lines as that of the equation for the 2-ppoint function.
The flow equation for $\Gamma_\kappa^{(n)}$ involves all $m$-point
functions with $m\le n+2$. Since the r.h.s. of the flow equations
have the structure of  one loop integrals, the  contributions
involving $\Gamma_\kappa^{(n+1)}$ and $\Gamma_\kappa^{(n+2)}$
  are of the types shown in Fig.~\ref{n-point-diagrams}. When the loop momentum
   entering these vertices  is taken to be zero, in line with eq.~(\ref{approx}), the vertex of order $n+1$  has one leg at zero
 momentum and the vertex of order $n+2$ has two; thus, according to eq.~(\ref{faireq=0}), they can be written as derivatives of the $n$-point function $\Gamma_\kappa^{(n)}$. It follows that
 the equation for $\Gamma_\kappa^{(n)}$ is a closed equation,
assuming of course that all other functions $\Gamma_\kappa^{(m)}$
with $m<n$ are determined similarly.

One can go beyond the leading order approximation, based on
eq.~(\ref{approx}), in the following way. Focusing on
eq.~(\ref{gamma2}) for  $\Gamma_\kappa^{(2)}(p,-p;\phi)$ one can solve  simultaneously the flow equations of
$\Gamma_\kappa^{(2)}$, $\Gamma_\kappa^{(3)}$ and $\Gamma_\kappa^{(4)}$, with no
approximation in the flow equation for $\Gamma_\kappa^{(2)}$, but using
(\ref{approx}) in the right-hand-side of the  flow equations for
$\Gamma_\kappa^{(3)}$ and $\Gamma_\kappa^{(4)}$. In this way one would determine
$\Gamma_\kappa^{(3)}$ and $\Gamma_\kappa^{(4)}$ with a ``leading order"
precision and $\Gamma_\kappa^{(2)}$ with a ``next-to-leading order" one.
By iterating the procedure, which amounts to including more equations, one would get  better approximations
for a larger number of $n$-point functions. It is easily verified that such an iterative   approximation scheme
reproduces
perturbation theory at high momenta: for instance, if the bare
vertices of the theory are momentum independent, the leading order
approximation contains the exact  one loop (this is only true if we keep
$q$ in the propagators, as we did in (\ref{gamma2app})). More generally, 
a simple analysis shows that to get the expression of a given
$n$-point functions at order $m$ loop, one has to consider all
equations  up to that for the $n+2m-2$-point function. (In a theory with
momentum dependent bare vertices, perturbation theory is also
reproduced, but more flow equations are required to match
perturbation theory at a given number of loops.) Of course, such a scheme is not only accurate at high momenta, where it reproduces perturbation theory, but also in the small momentum, possibly critical, regime, where its accuracy  is at least comparable
to that of the derivative expansion. In practice, the scheme that we have described  may become rapidly cumbersome. However, it can be simplified by using a further approximation consisting in a systematic expansion in $q^2/\kappa^2 $  
of the $n$-point functions in the right-hand-side of flow equations. This further approximation preserves the essential property of leading to a closed set  of equations. We have solved the flow equation for the 2-point function in leading order and obtained results which are comparable to those obtained in \cite{Blaizot:2004qa}. More complete  numerical studies  will   be presented in a forthcoming publication.

\section{The $O(N)$ model in the large $N$ limit}

We turn now to the $O(N)$ model in the large $N$ limit, where analytical results can be obtained. Our goal here is twofold. First,
to prove that, in its leading order, our approximation is in fact
exact in the large $N$ limit. Second, to solve analytically the
NPRG equations and find the $n$-point functions in a direct way,
i.e., without introducing an auxiliary field, as commonly done
\cite{Moshe:2003xn}.

The large $N$ limit of the $O(N)$ model within the NPRG has been
thoroughly  studied in \cite{D'Attanasio97}, where it is shown
that the effective action takes the form:
\begin{equation}\label{largeNlimit} \Gamma_\kappa[\phi]=\int d^dx
\left\{\frac{1}{2}\partial_{\mu}\phi^a\partial_{\mu}\phi^a\right\}+\hat\Gamma_\kappa[\rho],
\end{equation} where $\rho(x)=\phi^a(x)\phi^a(x)/2$, $a=1,\dots , N$. Using
this expression, the authors of Ref.~\cite{D'Attanasio97} show
that the LPA flow equation for the effective potential is exact.
Moreover, they find an interesting analytical solution of the
equation. However, the use of (\ref{largeNlimit}) in the flow
equation of a $n$-point function at zero field does not give a
closed equation \cite{Blaizot05}. In this section we obtain a
system of closed equations that can be solved analytically.

We shall again focus, for the sake of illustration, on the 2-point
function. Its exact flow equation in an external field is  the
generalization  of eq.~(\ref{gamma2})) to the $O(N)$ model:
\begin{eqnarray}
\label{gamma2champnonnul}
\partial_\kappa\Gamma_{ab}^{(2)}(p,-p;\kappa;\phi)&=&\int
\frac{d^dq}{(2\pi)^d}\partial_\kappa R_\kappa(q^2)\left\{G_{ij}(q^2;\kappa;\phi)\Gamma_{ajk}^{(3)}(p,q,-p-q;\kappa;\phi)\right. \nonumber \\
&&\times G_{kl}((q+p)^2;\kappa;\phi)\Gamma_{blm}^{(3)}(-p,p+q,-q;\kappa;\phi)G_{mi}(q^2;\kappa;\phi) \nonumber \\
&&\left.-\frac{1}{2}G_{ij}(q^2;\kappa;\phi)\Gamma_{abjk}^{(4)}(p,-p,q,-q;\kappa;\phi)G_{ki}(q^2;\kappa;\phi)\right\}.
\end{eqnarray}

At this point, it is useful to recall  the structure of the $n$-point functions
of the $O(N)$ model in the large N limit. These are obtained by taking the functional
derivatives of (\ref{largeNlimit}), letting the background field
to be constant, and then taking the Fourier transform. We obtain in this way
the following expressions for the 2-, 3- and 4-point vertices:
\beq \label{gamma2Ngrand}
\Gamma_{ab}^{(2)}(p,-p;\kappa;\phi)=(p^2+V'_\kappa
(\rho))\delta_{ab}+\phi_a\phi_b\hat{\Gamma}^{(2)}_\kappa(p,-p;\rho), \eeq
\beq
\Gamma_{abc}^{(3)}(p_1,p_2,p_3;\kappa;\phi)&=&\phi_a\delta_{bc}\hat{\Gamma}_\kappa^{(2)}(p_1,-p_1;\rho)+\phi_b\delta_{ac}\hat{\Gamma}_\kappa^{(2)}(p_2,-p_2;\rho)\nonumber \\
&+&\phi_c\delta_{ab}\hat{\Gamma}_\kappa^{(2)}(p_3,-p_3;\rho)
+\phi_a\phi_b\phi_c\hat{\Gamma}_\kappa^{(3)}(p_1,p_2,p_3;\rho),
\eeq
\beq\label{Gamma4}
&&\Gamma_{abcd}^{(4)}(p_1,p_2,p_3,p_4;\kappa;\phi)=\delta_{ab}\delta_{cd}\hat{\Gamma}_\kappa^{(2)}(p_1+p_2,-p_1-p_2;\rho)\nonumber \\
&&+\delta_{ac}\delta_{bd}\hat{\Gamma}_\kappa^{(2)}(p_1+p_3,-p_1-p_3;\rho)
+\delta_{ad}\delta_{bc}\hat{\Gamma}_\kappa^{(2)}(p_1+p_4,-p_1-p_4;\rho)\nonumber \\
&&+\delta_{ab}\phi_c\phi_d\hat{\Gamma}_\kappa^{(3)}(p_1+p_2,p_3,p_4;\rho)
+\delta_{ac}\phi_b\phi_d\hat{\Gamma}_\kappa^{(3)}(p_1+p_3,p_2,p_4;\rho)\nonumber \\
&&+\delta_{ad}\phi_b\phi_c\hat{\Gamma}_\kappa^{(3)}(p_1+p_4,p_2,p_3;\rho)
+\delta_{bc}\phi_a\phi_d\hat{\Gamma}_\kappa^{(3)}(p_2+p_3,p_1,p_4;\rho) \nonumber \\
&&+\delta_{bd}\phi_a\phi_c\hat{\Gamma}_\kappa^{(3)}(p_2+p_4,p_1,p_3;\rho)
+\delta_{cd}\phi_a\phi_b\hat{\Gamma}_\kappa^{(3)}(p_3+p_4,p_1,p_2;\rho)\nonumber \\
&&+\phi_a\phi_b\phi_c\phi_d\hat{\Gamma}_\kappa^{(4)}(p_1,p_2,p_3,p_4;\rho) \; .
\eeq
As for the propagator, it  can be written in terms of its
longitudinal and transverse components:
\begin{equation}
G_{ab}(p^2;\phi)=G_T(p^2;\rho)\left(\delta_{ab}-\frac{\phi_a\phi_b}{2\rho}\right)
+G_L(p^2;\rho)\frac{\phi_a\phi_b}{2\rho},
\end{equation}
with
\begin{eqnarray}
\label{propagatorsTL}
G_T^{-1}(p^2;\kappa;\rho)&=&p^2+V_\kappa'(\rho)+R_\kappa(p^2)\;,\nonumber \\
G_L^{-1}(p^2;\kappa;\rho)&=&p^2+V_\kappa'(\rho)+2\rho \hat{\Gamma}_\kappa^{(2)}(p,-p;\rho) +R_\kappa(p^2).
\end{eqnarray}

We can now replace these expressions in
eq.~(\ref{gamma2champnonnul}) and keep only the leading terms in
$1/N$. To do so, notice that the
 non trivial large $N$ limit of the $O(N)$ model is obtained when
the effective action $\Gamma_\kappa$ is considered to be of order
$N$ and the $\phi$ field of order $\sqrt{N}$ (see for example
\cite{D'Attanasio97}). Then, the 2-point function is of order 1,
the 3-point function of order $1/\sqrt{N}$ and the 4-point
function of order $1/N$. Simple counting of $N$ factors in the
flow equation (\ref{gamma2champnonnul}) gives a right hand side of order $1/N$. Thus, the
only way of having both sides of order 1 is to collect an explicit
$N$ factor from the trace of an identity matrix. The
surviving terms yield:
\begin{eqnarray}
\label{gamma2Ngrand+}
&&\partial_\kappa\Gamma_{ab}^{(2)}(p,-p;\kappa;\phi)=N\int
\frac{d^dq}{(2\pi)^d}\partial_\kappa R_\kappa(q^2)G_T^2(q^2;\kappa;\rho)
\left\{-\frac{1}{2}\delta_{ab}\hat{\Gamma}_\kappa^{(2)}(0,0;\rho)\right.\nonumber \\
&\qquad&\qquad\left.+\phi_a\phi_b\left(\hat{\Gamma}_\kappa^{(2)}(p,-p;\rho)\right)^2 G_T((p+q)^2;\kappa;\rho)
-\frac{1}{2}\phi_a\phi_b
\hat{\Gamma}_\kappa^{(3)}(p,-p,0;\rho)\right\}.
\end{eqnarray}

Now comes the important observation: the vertices contributing to
the flow in the large-$N$ limit are $q$-independent. That means
that our leading order approximation becomes exact in the
large-$N$ limit of the $O(N)$ model, as anticipated. Notice that
for this to be true,  we need to keep the $q$ dependence of the
propagators in the flow equations. However,  in the large-N limit, the momentum
dependence of the transverse propagator (the only one appearing in
(\ref{gamma2Ngrand+})) is simply the bare
one, as shown in eq.~(\ref{propagatorsTL}).

Substituting in eq.~(\ref{gamma2Ngrand+}) the expression (\ref{gamma2Ngrand}) of the 2-point function
 and performing the tensor decomposition,
 one obtains the system of
equations:
\begin{eqnarray}\label{closedeqnsON}
\partial_\kappa V'_\kappa(\rho)&=&-\frac{N}{2}V''_\kappa(\rho)\int \frac{d^dq}{(2\pi)^d}\partial_\kappa R_\kappa(q^2)G_T^2(q^2;\kappa;\rho) \nonumber \\
\partial_\kappa \hat{\Gamma}_\kappa^{(2)}(p,-p;\rho)&=&N\int \frac{d^dq}{(2\pi)^d}\partial_\kappa
R_\kappa(q^2)G_T^2(q^2;\kappa;\rho) \nonumber \\
&&\times\left\{\left(\hat{\Gamma}_\kappa^{(2)}(p,-p;\rho)\right)^2 G_T((p+q)^2;\kappa;\rho)
-\frac{1}{2} \frac{\partial \hat{\Gamma}_\kappa^{(2)}(p,-p;\rho)}{\partial \rho}\right\}
\end{eqnarray}
where we have used, as in eq.~(\ref{faireq=0}),
\begin{equation}
\hat{\Gamma}_\kappa^{(2)}(0,0;\rho)=V''_\kappa (\rho) ,\qquad
\hat{\Gamma}_\kappa^{(3)}(p,-p,0;\rho)=\frac{\partial \hat{\Gamma}_\kappa^{(2)}(p,-p;\rho)}{\partial \rho},
\end{equation}
which allowed us to close the equations. In the equations above, $V'_\kappa(\rho)$ and $V''_\kappa(\rho)$ denote respectively the first and second derivative of the effective potential with respect to $\rho$.

The first of eqs.~(\ref{closedeqnsON})  is simply the derivative of the equation for
the potential in the large-N limit which was solved analytically
using two different methods \cite{Tetradis95,D'Attanasio97}. The
second equation, which, to our knowledge, has not been presented
before in the literature, can be solved also  using any of these
two methods. Here we present the solution using the method of
\cite{Tetradis95}.

To this aim, we set $W=V'_\kappa (\rho)$, and use $(\kappa,W)$ as
independent variables instead of $(\kappa,\rho)$. In order to do
so, we introduce the inverse function relating $\rho$ to $W$:
\begin{equation}
\rho=f_\kappa(W) \end{equation} so that:
\begin{equation}\label{relations}
\partial_\kappa f_\kappa (W)=-f'_\kappa (W)\partial_\kappa W(\rho) ,\qquad
f'_\kappa (W)=1/W'(\rho),
\end{equation}
where $f'$ and $W'$ are the derivatives of  $f$ and $W$ with
respect to their respective arguments. Making the change of
variables in eq.~(\ref{closedeqnsON}), and using the relations
(\ref{relations}), one gets:
\begin{eqnarray}
\partial_\kappa f_\kappa(W)&=&\frac{N}{2}\int \frac{d^dq}{(2\pi)^d}\frac{\partial_\kappa R_\kappa (q^2)}{(q^2+W+R_\kappa (q^2))^2} \nonumber \\
\partial_\kappa \left(\hat{\Gamma}_\kappa^{(2)}(p,-p;\rho)\right)^{-1}
&=&-N\int \frac{d^dq}{(2\pi)^d}\frac{\partial_\kappa R_\kappa (q^2)}{(q^2+W+R_\kappa (q^2))^2}
\frac{1}{(q+p)^2+W+R_\kappa ((q+p)^2)} \nonumber \\
\end{eqnarray}
At this point we observe that the right hand sides of  both the
expressions above  are total derivatives:
\begin{eqnarray}
&&\int \frac{d^dq}{(2\pi)^d}\frac{\partial_\kappa R_\kappa (q^2)}{(q^2+W+R_\kappa (q^2))^2}
=- \partial_\kappa \int \frac{d^dq}{(2\pi)^d}\frac{1}{q^2+W+R_\kappa (q^2)}\nonumber \\
&&\int \frac{d^dq}{(2\pi)^d}\frac{\partial_\kappa R_\kappa (q^2)}{(q^2+W+R_\kappa (q^2))^2}\,\frac{1}{(q+p)^2+W+R_\kappa ((q+p)^2)} \nonumber \\
&&=-\frac{1}{2}\partial_\kappa \int \frac{d^dq}{(2\pi)^d}\frac{1}{q^2+W+R_\kappa (q^2)}\,\frac{1}{(q+p)^2+W+R_\kappa ((q+p)^2)}.\nonumber \\
\end{eqnarray}
This allows us to integrate analytically the equations. For the potential, one obtains the well known result \cite{Tetradis95,Berges00}.
By imposing the initial condition that for $\kappa =\Lambda$ the 4-point function is the bare vertex, we get from eq.~(\ref{Gamma4}) 
\begin{equation}
\hat{\Gamma}_{\kappa=\Lambda}^{(2)}(p,-p;\rho)=\frac{u}{3},
\end{equation}
 corresponding to a $\phi^4$ interaction in the classical action, of the form $(u/4!)( \phi_a\phi_a)^2$. We then obtain the
longitudinal part of the 2-point function (in the limit $\Lambda\to\infty$):
\begin{equation}
\hat{\Gamma}_\kappa^{(2)}(p,-p;\rho)=\frac{u}{3}
\frac{1}{1+\frac{Nu}{6}\int \frac{d^dq}{(2\pi)^d}\frac{1}{q^2+W+R_\kappa (q^2)}\frac{1}{(q+p)^2+W+R_\kappa ((q+p)^2)}}.
\end{equation}
This is the usual sum of bubbles in large $N$, which can be found using other methods \cite{Moshe:2003xn}. Notice however that our solution follows directly from the NPRG without introducing any auxiliary field. The dependence on $W$ reflects the presence of the background field: $W$ plays here the role of a (field dependent)  ``mass term''.

The same method can be applied to the other $n$-point functions.
The calculations are straightforward, but the explicit results
involve lengthy expressions. Let us just give here the result for
the 3-point function: \beq
\hat\Gamma^{(3)}_\kappa(p_1,p_2,p_3;W) &=&N\hat\Gamma^{(2)}_\kappa(p_1,-p_1;W) \hat\Gamma^{(2)}_\kappa(p_2,-p_2;W) \hat\Gamma^{(2)}_\kappa(p_3,-p_3;W)
\nonumber \\
&&\times\int \frac{d^dq}{(2\pi)^d} G_T(q^2;\kappa;\rho)
G_T((q+p_1)^2;\kappa;\rho)G_T((q-p_3)^2;\kappa;\rho) .\nonumber \\
\eeq

\section{Conclusions and perspectives}

In this paper we have presented an approximation scheme to solve
the NPRG equations and obtain the
$n$-point functions  for any external momenta. 
Our proposal is as general and systematic as the
derivative expansion: results can be systematically improved as
one can write, at any order, a closed set of equations.
We have
provided here, as an illustration, an application to the
$O(N)$ model  in the
large $N$ limit: there, the method  turns out to be exact at leading order, 
and it provides an economical  procedure to obtain the analytical  expressions  
of the $n$-point functions at arbitrary momenta. Clearly, however, the method is more general. It is based on
an approximation that can be  exported to other field theories. 
Work on such extensions, in particular to gauge theories, is
underway. This, as well as numerical studies for physically
interesting problems, will be reported in forthcoming
publications.

\bibliographystyle{unsrt}

\end{document}